\documentclass[pra,aps,superbib,citeautoscript,twocolumn]{revtex4-1}

\usepackage[colorlinks=true,urlcolor=blue,citecolor=blue,linkcolor=blue]{hyperref}

\usepackage{amsmath,bm}
\usepackage{amstext}
\usepackage{epsfig}
\usepackage{xcolor}
\usepackage{subfig}
\usepackage{graphicx,amsmath,amssymb,tabularx}
\usepackage{multirow}
\usepackage{array}
\usepackage{dsfont}
\usepackage{caption}
\usepackage{cleveref}
\usepackage{ulem}
\usepackage[toc]{appendix}
\usepackage{physics}
\usepackage{mathrsfs}  
\usepackage{amsbsy}

\definecolor{dgreen}{rgb}{0,.5,0}
\definecolor{dred}{rgb}{.7,.0,.0}

\DeclareMathAlphabet\mathbfcal{OMS}{cmsy}{b}{n}

\newcommand{\ie}{{\it i.e.}}

\newcommand{\manu}[1]{{\textcolor{blue}{Manu: #1}} }

\newcommand{\be}{\begin{eqnarray}}
\newcommand{\ee}{\end{eqnarray}}
\DeclareMathAlphabet\mathbfcal{OMS}{cmsy}{b}{n}

\newcommand{\bgam}{\bm{\gamma}}
\newcommand{\bV}{\bm{V}}
\newcommand{\vac}{\rm{vac}}
\newcommand{\rev}[1]{{{#1}}}

\begin{document}

\title{
A unified density matrix functional construction of quantum baths in density matrix embedding theory beyond the mean-field approximation
}

\author{
Sajanthan Sekaran$^1$, Oussama Bindech$^1$, and Emmanuel Fromager$^1$
}
\affiliation{\it 
~\\
$^1$Laboratoire de Chimie Quantique,\\
Institut de Chimie, CNRS / Universit\'{e} de Strasbourg,\\
4 rue Blaise Pascal, 67000 Strasbourg, France\\
}


\begin{abstract}

The equivalence in \rev{one-electron} quantum bath between \rev{the practical implementation of} density matrix embedding theory (DMET) and the more recent Householder-transformed density matrix functional embedding theory has been shown previously in the standard but special case where the reference full-size (one-electron reduced) density matrix, from which the bath is constructed, is idempotent [J. Chem. Phys. 157, 214112 (2022)]. We prove mathematically that the equivalence remains valid when the density matrix is not idempotent anymore, thus allowing for the construction of correlated (one-electron) quantum baths. A density-matrix functional exactification of DMET is derived within the present unified quantum embedding formalism. Numerical examples reveal that the embedding cluster can be quite sensitive to the level of density-matrix functional approximation used for computing the reference density matrix.     
\end{abstract}

\maketitle



\section{Introduction}

Quantum embedding is an electronic structure calculation strategy that relies on the fragmentation of the (molecular or extended) system  under study~\cite{sun2016quantum,IJQC20_Adam-Michele_embedding_special_issue}. When performed on a lattice, in real space, or in a localized orbital basis, such a fragmentation allows for an efficient description of strong local electronic correlations. Obviously, the Schrödinger equation or its reformulation in terms of a simpler quantity than the wave function, like the one-particle Green's function, for example, cannot be solved for each fragment individually. The reason is that a given fragment is entangled with its environment, which consists of the remaining fragments if all of them are disjoint. Providing a mathematical simplification (usually referred to as quantum bath) of the environment that can be used for describing the fragment as part of a bigger system is a key aspect of quantum embedding theory.\\

{\it Density matrix embedding theory} (DMET)~\cite{knizia2012density,wouters2016practical,wu2019projected,Wu2020Enhancing,Li2023Multi-site} has emerged over the last decade as a promising numerical approach to strongly correlated electrons in both quantum chemistry and condensed matter physics for several reasons. First of all, unlike the well-established {\it dynamical mean-field theory} (DMFT)~\cite{georges1992hubbard,georges1996limitdimension,kotliar2004strongly,held2007electronic,zgid2011DMFTquantum} of extended systems, DMET uses a drastically simplified one-electron bath that contains as many orbitals as the embedded fragment. As a result, the fragment$+$bath system (so-called embedding cluster) can be solved accurately (if not exactly) by means of wave function-based methods, thus allowing for the application of quantum chemical methods to strongly correlated materials. Its application and development in the context of quantum chemistry is also an active field of research~\cite{welborn2016bootstrap,wouters2016practical,ye2018incremental,hermes2019multiconfigurational,hermes2020variational,nusspickel2023effective}. Extensions to non-equilibrium electron dynamics~\cite{kretchmer2018real} and finite temperatures~\cite{Sun2020Finite-temperature} have also been explored more recently. Let us mention that  alternative embedding strategies, which are close in spirit to DMET but rely on a different formalism (namely the exact factorization of the electronic wave function), have been proposed recently~\cite{lacombe2020embedding,Requist2021Fock-Space}.\\

At a more fundamental level, formal connections with DMFT~\cite{fertitta2018rigorous,JCP19_Booth_Ew-DMET_hydrogen_chain,PRB21_Booth_effective_dynamics_static_embedding}, the 
{\it rotationally-invariant slave-boson} (RISB) method~\cite{ayral2017dynamical,lee2019rotationally,PRX21_Lee_SlaveBoson_resp_functions-superconductivity,Lee2023Accuracy}, and the ghost Gutzwiller approximation~\cite{lanata2023derivation} have been established. These works were originally motivated by the \rev{ill-conditioned} self-consistency loop of practical DMET calculations which aims at mapping the fragment block of a correlated \rev{\it pure-state} (and therefore non-idempotent) embedding cluster's one-electron reduced density matrix (1RDM) onto the fragment block of a full-size \rev{\it pure-state} and mean-field-like (and therefore idempotent) 1RDM~\cite{lee2019rotationally,Faulstich2022Pure}. The latter constraint was originally proposed by analogy with DMFT and its mapping constraint of the local one-particle Green's function~\cite{knizia2012density}. We refer the reader to Ref.~\cite{cances2023mathematical} for a mathematical insight into DMET and, in particular, into the self-consistency loop through first and second orders in the two-electron interaction strength. Note that, if the mapping is restricted to the density, {\it i.e.}, the diagonal elements of the 1RDM, like in {\it density embedding theory} (DET)~\cite{bulik2014density}, the convergence properties of the embedding algorithm are greatly improved. More recently, connections with {\it Kohn--Sham density functional theory} (KS-DFT) have been established and exploited in self-consistent density embedding (SDE) calculations~\cite{mordovina2019self}. A density-functional exactification of DET has also been derived in Ref.~\onlinecite{sekaran2022local} for the Hubbard model, thus leading to an alternative {\it local potential functional embedding theory} (LPFET)~\cite{sekaran2022local}. Nevertheless, involving off-diagonal elements of the 1RDM into the self-consistent construction of the bath enables to capture more physics~\cite{knizia2012density}. As briefly mentioned in Ref.~\onlinecite{sekaran2021} , one could reconsider the bath as a functional of the 1RDM (that we refer to as the density matrix in the rest of the paper, for simplicity), thus allowing for the construction of a correlated quantum bath through the use of a non-idempotent full-size \rev{one-electron reduced} density matrix. We may also benefit from the latest developments in {\it natural orbital functional theory} (NOFT) ~\cite{Piris2017Global,Piris2021Global,Huan2023Outstanding,Lew2023Electron} for computing and exploiting full-size correlated density matrices. The more recent {\it Householder-transformed density matrix functional embedding theory} (Ht-DMFET)~\cite{sekaran2021,Yalouz2022quantum_embedding} was originally developed independently in this spirit.\\

Even though the mathematical construction of the bath in Ht-DMFET differs substantially from that of standard DMET, it has been shown very recently that they are actually equivalent in the usual but special case of a non-interacting or mean-field ({\ie}, idempotent) full-size density matrix [see Appendix D of Ref.~\onlinecite{Yalouz2022quantum_embedding}]. The proof is based on the regular DMET construction of the bath from the overlap matrix between the occupied molecular orbitals and the fragment orbitals~\cite{wouters2016practical}, which of course does not extend trivially to correlated density matrices where possibly all (natural) molecular orbitals are (fractionally) occupied. We mathematically prove in this work that the equivalence in (one-electron) quantum bath between Ht-DMFET and DMET holds even if the reference full-size density matrix is not idempotent anymore. As a direct and important consequence of our finding, we show that a density-matrix functional exactification of DMET \rev{(where the one-electron quantum bath becomes a functional of the true correlated ground-state full-size density matrix)} can be derived, thus paving the way towards a systematic improvement of the embedding solely based on the (static) density matrix. The paper is organized as follows. We briefly introduce in Sec.~\ref{sec:motivations_notations} the motivations, philosophy, and terminology of DMET as well as the notations used throughout the paper. The mathematical constructions of DMET and Ht-DMFET quantum baths are reformulated in Secs.~\ref{sec:SVD_EF_block} and \ref{sec:Householder}, respectively, in terms of a global unitary transformation in the one-electron Hilbert space, for ease of comparison. The key result of the paper, which is the equivalence of the two baths beyond the full-size mean-field level of calculation, is proved in Sec.~\ref{sec:equivalence_SVD-HH}. The formal exactification of the embedding procedure is discussed in Sec.~\ref{sec:exactification_dmet}, followed by illustrative numerical results obtained for Hubbard rings (see Sec.~\ref{sec:numerical_examples}). Conclusions and perspectives are finally given in Sec.~\ref{sec:conclusions}.

\section{Motivation and notations}\label{sec:motivations_notations}

Let $\left\{\ket{\chi_i}=\hat{c}^\dagger_i\ket{\rm vac}\right\}_{1\leq i\leq {L}_{\rm tot}}$ be an orthonormal basis of a given one-electron spin-orbital space with total dimension $L_{\rm tot}$ (we use second-quantized notations for convenience). The latter obviously depends on the system for which we want to solve the (ground-state) electronic Schr\"{o}dinger equation $\hat{H}\ket{\Psi_0}=E_0\ket{\Psi_0}$, where the full-system Hamiltonian contains one- ($\hat{h}$) and two-electron ($\hat{W}$) terms:
\be\label{eq:full_system_Hamil}
\begin{split}
\hat{H}&=\hat{h}+\hat{W}
\\
&\equiv\sum^{{L}_{\rm tot}}_{i,j=1}h_{ij}\hat{c}^\dagger_i\hat{c}_j+\dfrac{1}{2}\sum^{{L}_{\rm tot}}_{i,j,k,l=1}\langle ij\vert kl \rangle \hat{c}^\dagger_i\hat{c}^\dagger_j \hat{c}_l \hat{c}_k.
\end{split}
\ee
In practical embedding calculations, \rev{the one-electron basis $\{i,j,k,l,\ldots\}$} usually consists of {\it localized} (in real space) spin-orbitals, thus allowing for a fragmentation of the molecule or the extended system under study \rev{and a subsequent adequate description of strong local electron correlations.} \rev{Obviously, the localization procedure (of canonical Hartree--Fock molecular orbitals, for example) is necessary in quantum chemical calculations since the bare atomic orbitals overlap and, therefore, are not orthogonal}. In the particular case of a lattice model such as the single-orbital Hubbard model, the indices $i,j,k,l$ refer to atomic sites (strictly speaking, they refer to one-electron atomic spin states) and ${L}_{\rm tot}=2L$, where $L$ is the number of sites in the lattice. The localized spin-orbital representation \rev{used in quantum chemistry} is sometimes referred to as the lattice representation for that reason. A to-be-embedded {\it fragment} is arbitrarily designed in this context by selecting ${L}_{\rm frag}$ localized spin-orbitals. The latter are usually referred to as impurities once they have been embedded into a (traditionally but not compulsorily non-interacting) quantum bath~\cite{wouters2016practical}. 
We should stress at this point that the self-consistent optimization of the bath (that relies on density matrix elements mapping constraints) and how to optimally combine several embedded fragments in the computation of total (molecular) energies are separate (and still open~\cite{nusspickel2023effective}) issues that will not be addressed here. The purpose of the present work is to provide a unified description of one-electron quantum baths that are employed in quantum embedding computations. Our starting point will be the following decomposition of the full one-electron Hilbert space into fragment and environment spin-orbitals,  
\be\label{eq:decomp_fragment_env}
\left\{\ket{\chi_i}\right\}=\underset{fragment}{\underbrace{\left\{\ket{\chi_f}\right\}_{1\leq f\leq {L}_{\rm frag}}}}\oplus
\underset{environment}{\underbrace{\left\{\ket{\chi_e}\right\}_{{L}_{\rm frag}<e\leq {L}_{\rm tot}}}}.
\ee
In practical DMET calculations, designing a quantum bath usually consists in extracting from the one-electron fragment's environment subspace a limited number (usually ${L}_{\rm frag}$, for reasons that will become clearer later on) of spin-orbitals $\left\{\phi_b\right\}$ that form with the fragment spin-orbitals an active spin-orbital space (referred to as the {\it cluster} spin-orbital space) in which an integer number (equal to ${L}_{\rm frag}$, for reasons that will also become clearer in Sec.~\ref{sec:exact_embedding_idempotRDM}) of electrons can be distributed, \rev{in analogy} with the {\it complete active space configuration interaction} (CASCI) method~\cite{pinkbible_CI_chapter}. As a result, the full one-electron space is now further decomposed as follows,
\be\label{eq:decomp_fragment_bath_cenv}
\left\{\ket{\chi_i}\right\}=\overset{cluster}{\overbrace{
\underset{fragment}{\underbrace{\left\{\ket{\chi_f}\right\}}}
\oplus
\underset{bath}{\underbrace{\left\{\ket{\phi_b}\right\}}}}}
\oplus
\overset{cluster's\,environment}{\overbrace{\left\{\ket{\phi_{\mathcal{E}}}\right\}}},
\ee
where ${L}_{\rm frag}<b\leq 2{L}_{\rm frag}$ and $2{L}_{\rm frag}<\mathcal{E}\leq {L}_{\rm tot}$, and the ${L}_{\rm frag}$-electron Schr\"{o}dinger equation is solved for the cluster. \rev{Note that, unlike in conventional CASCI calculations, where the active orbital space consists of {\it delocalized} molecular orbitals, the cluster orbital space (which plays the role of an active orbital space) consists of the {\it localized} fragment orbitals plus the bath orbitals (that are delocalized over the fragment's environment).} Note also that the (so-called {\it core}) electrons that remain inactive in the embedding calculation fully occupy spin-orbitals that belong to the embedding cluster's environment subspace $\left\{\phi_{\mathcal{E}}\right\}$. Situations where the cluster is exactly disentangled from its environment will be discussed in Sec.~\ref{sec:exact_embedding_idempotRDM}. Note that we are ultimately interested in the properties of the embedded fragment. The full ``fragment+bath'' embedding cluster, in which the bath plays the role of an electronic reservoir, \rev{is a compact simplification of the true physical system from which local (fragment) properties can be extracted. The embedding cluster's Hamiltonian is obtained by projecting the true Hamiltonian (or parts of it, if a non-interacting bath is employed) onto the ``fragment+bath'' subspace (see Sec.~\ref{sec:exact_interacting_embedding}) and provides, in this respect, a physical approximation to the true system.} 

\section{Quantum bath from the SVD of the environment-fragment density matrix block}\label{sec:SVD_EF_block}
In order to prove that the bath spin-orbital subspaces generated from DMET and the more recent Ht-DMFET are identical, we revisit in the following one of the numerous~\cite{zheng2018density,Wu2020Enhancing} formulations of DMET where the bath is constructed directly from the density matrix of the full system (written in the original localized spin-orbital basis)~\cite{Li2023Multi-site}:
\be\label{eq:1RDM_full-size_elements_lattice_rep}
\bgam\equiv
\left\{\gamma_{ij}=\expval{\hat{c}^\dagger_i\hat{c}_j}\right\}_{1\leq i,j\leq L_{\rm tot}}.
\ee  
At this point we do not specify if $\bgam$ is exact, approximately correlated or not correlated at all. That statement implies that we do not exclude the possibility to construct a (one-electron) quantum bath that is correlated through the density matrix, even though this strategy is never adopted in conventional DMET calculations. Indeed, in the latter case, a (idempotent) mean-field-like density matrix is employed, for convenience. Let us stress that, even if $\bgam$ were the exact ground-state full-size density matrix, the resulting correlated (one-electron) bath, as described in the following, would {\it not} be exact in the sense that solving the Schr\"{o}dinger equation for the closed ``fragment+bath'' subsystem would not provide an exact description of the fragment properties~\cite{sekaran2021}. The true bath consists of {\it many-body} wave functions that are constructed from the Schmidt decomposition of the exact ground-state many-body wave function~\cite{knizia2012density}. Nevertheless, as discussed in further details in Sec.~\ref{sec:exact_interacting_embedding}, combining a density matrix functional one-electron quantum bath with the proper complementary correlation density matrix functional can formally exactify the embedding.\\

Returning to Eq.~(\ref{eq:1RDM_full-size_elements_lattice_rep}), we decompose the density matrix into blocks as follows, according to Eq.~(\ref{eq:decomp_fragment_env}),
\be\label{eq:1RDM_local_basis}
\bgam=
\left[\begin{matrix}
\bgam_{ff} & \bgam^\dagger_{ef}
\\
\bgam_{ef}& \bgam_{ee}
\end{matrix}
\right].
\ee
Note that the indices $f$ and $e$ in Eq.~(\ref{eq:1RDM_local_basis}) do not refer to specific matrix elements. They have been introduced in order to easily identify the matrix blocks (in bold) and their dimensions. For example, the environment-fragment block reads   
$\bgam_{ef}\equiv \left\{\gamma_{ij}\right\}_{{L}_{\rm frag}<i\leq {L}_{\rm tot},1\leq j\leq {L}_{\rm frag}}$. The fact that it is non-zero obviously prevents us from treating the fragment as a separate subsystem. Nevertheless, we can identify a one-electron subspace [which corresponds to the bath introduced in Eq.~(\ref{eq:decomp_fragment_bath_cenv})] to which the fragment will, ultimately, be exclusively entangled. For that purpose, we consider the {\it singular value decomposition} (SVD) of $\bgam_{{e}f}$~\cite{Li2023Multi-site},
\be\label{eq:SVD_ef_block_1RDM}
\bgam_{{e}f}={\bm{U}}_{ee}{\bm\sigma}_{eb}{\bm{V}}^\dagger_{fb},
\ee
where
\be
{\bm\sigma}_{eb}=
\left[
\begin{matrix}
\mathbfcal{D}^{\frac{1}{2}}
\\
{\bm 0}_{\mathcal{E}b}
\end{matrix}
\right],
\ee
and 
\be\label{eq:non-zero_square_singular_values_matrix}
\mathbfcal{D}\equiv\left\{\delta_{bb'}\sigma^2_b\right\}
\ee 
is the ${L}_{\rm frag}\times{L}_{\rm frag}$ diagonal matrix of the square singular values that is obtained by diagonalizing the Hermitian matrix 
$\bgam_{{e}f}^\dagger\bgam_{{e}f}$:
\be\label{eq:diag_pb_SVD}
\mathbfcal{U}^\dagger\bgam_{{e}f}^\dagger\bgam_{{e}f}\mathbfcal{U}=\mathbfcal{D},
\ee
$\mathbfcal{U}^\dagger\mathbfcal{U}=\mathbfcal{U}\mathbfcal{U}^\dagger={\bm
1}_{ff}$ being the ${L}_{\rm frag}\times{L}_{\rm frag}$ identity matrix. We implicitly assumed in Eq.~(\ref{eq:SVD_ef_block_1RDM}) that the dimension of the fragment is smaller than that of its environment, {\ie}, ${L}_{\rm frag}<{L}_{\rm tot}-{L}_{\rm frag}$ or, equivalently, ${L}_{\rm frag}<\frac{{L}_{\rm tot}}{2}$. This is always the case in practical calculations where ${L}_{\rm frag}$ is taken as small as possible in order to reduce the computational cost of the embedding calculation. Moreover, $\bgam_{{e}f}$ is assumed to be a full rank matrix (as observed in practical calculations), which implies that the dimension of the bath equals $\rank(\bgam_{{e}f})={L}_{\rm frag}$. It also implies that all the singular values in Eq.~(\ref{eq:non-zero_square_singular_values_matrix}) differ from zero. Indeed, if there exists an eigenvector $\underline{\mathcal{U}}$ of $\bgam_{{e}f}^\dagger\bgam_{{e}f}$ associated to the eigenvalue zero then   
\be
\left(\bgam_{{e}f}\,\underline{\mathcal{U}}\right)^\dagger\bgam_{{e}f}\,\underline{\mathcal{U}}=\underline{\mathcal{U}}^\dagger\bgam_{{e}f}^\dagger\bgam_{{e}f}\,\underline{\mathcal{U}}=0,
\ee
or, equivalently, $\bgam_{{e}f}\,\underline{\mathcal{U}}=\underline{0}$, thus leading to $\underline{\mathcal{U}}=\underline{0}$ (since $\bgam_{{e}f}$ is full rank), which is of course impossible. As a result, $\mathbfcal{D}$ is {\it invertible} and
\be\label{eq:inverse_diag_sqSV_matrix}
\mathbfcal{D}^{-1}\equiv\left\{\dfrac{\delta_{bb'}}{\sigma^2_b}\right\}.
\ee
\\
Returning to the SVD of $\bgam_{{e}f}$, the left unitary matrix ${\bm{U}}_{ee}$ in Eq.~(\ref{eq:SVD_ef_block_1RDM}) can be split into bath and cluster's environment blocks [see Eq.~(\ref{eq:decomp_fragment_bath_cenv})],   
\be\label{eq:decomp_Uee_unitary_transf}
{\bm{U}}_{ee}=
\left[\begin{matrix}
{\bm{U}}_{eb} & {\bm{U}}_{e\mathcal{E}}\\
\end{matrix}
\right],
\ee 
where a complete set of orthonormal $({L}_{\rm tot}-{L}_{\rm frag})$-row column vectors [$({L}_{\rm tot}-2{L}_{\rm frag})$ vectors in total] that is orthogonal to the columns of $\bgam_{{e}f}$ are collected in ${\bm{U}}_{e\mathcal{E}}$, \ie, 
\be\label{eq:SVD_orthonormal_clustersenv}
{\bm{U}}_{e\mathcal{E}}^\dagger {\bm{U}}_{e\mathcal{E}}={\bm 1}_{\mathcal{E}\mathcal{E}}
\ee
and
\be\label{eq:orthogonality_constraint_mathcalE}
\bgam^\dagger_{{e}f}{\bm{U}}_{e\mathcal{E}}={\bm 0}_{f\mathcal{E}}.
\ee
The remaining ${L}_{\rm frag}$ orthonormal bath column vectors are collected in $\bm{U}_{eb}$, which reads
\be\label{eq:SVD_U1matrix}
{\bm U}_{eb}=\bgam_{{e}f}\mathbfcal{U}{\mathbfcal{D}}^{-\frac{1}{2}}.
\ee
Finally, the right unitary matrix in Eq.~(\ref{eq:SVD_ef_block_1RDM}) is determined from ${\bm
U}_{eb}$ as follows,
\be
{\bm{V}}_{fb}=\bgam_{{e}f}^\dagger{\bm
U}_{eb}{\mathbfcal{D}}^{-\frac{1}{2}},
\ee
which, according to Eqs.~(\ref{eq:diag_pb_SVD}) and (\ref{eq:SVD_U1matrix}), simplifies to
\be
{\bm{V}}_{fb}=\mathbfcal{U}.
\ee
Note that ${\bm{U}}_{ee}$ is unitary by construction. Indeed, according to Eqs.~(\ref{eq:diag_pb_SVD}), (\ref{eq:SVD_orthonormal_clustersenv}), (\ref{eq:orthogonality_constraint_mathcalE}), and (\ref{eq:SVD_U1matrix}),
\be
{\bm{U}}_{ee}^\dagger{\bm{U}}_{ee}
= 
\left[
\begin{matrix}
{\bm U}_{eb}^\dagger {\bm U}_{eb} & {\bm U}_{eb}^\dagger {\bm U}_{e\mathcal{E}}
\\
{\bm U}_{e\mathcal{E}}^\dagger {\bm U}_{eb}
&
{\bm U}_{e\mathcal{E}}^\dagger {\bm U}_{e\mathcal{E}}
\end{matrix}
\right]={\bm 1}_{ee},
\ee
thus leading to ${\bm{U}}_{ee}^\dagger={\bm{U}}_{ee}^{-1}$ and, therefore,
\be\label{eq:RI_SVD}
{\bm{U}}_{ee}{\bm{U}}_{ee}^\dagger={\bm U}_{eb}{\bm U}_{eb}^\dagger+{\bm U}_{e\mathcal{E}}{\bm U}_{e\mathcal{E}}^\dagger={\bm 1}_{ee}.
\ee

Now that we have identified all the ingredients of the SVD in Eq.~(\ref{eq:SVD_ef_block_1RDM}), we can apply the following unitary transformation to the full-system density matrix,
\be
\bgam\rightarrow {\bm U}^\dagger\bgam{\bm U},
\ee
where 
\be\label{eq:full_unitary_transf_matrix}
{\bm U}=
\left[
\begin{matrix}
{\bm
1}_{ff} & {\bm 0}_{fe}\\
 {\bm 0}_{ef} & {\bm{U}}_{ee}
 \end{matrix}
\right]
\ee
is a functional of $\bgam_{{e}f}$, as readily seen from Eqs.~(\ref{eq:diag_pb_SVD}), (\ref{eq:decomp_Uee_unitary_transf}), (\ref{eq:orthogonality_constraint_mathcalE}), and (\ref{eq:SVD_U1matrix}), which is equivalent to performing the following change in spin orbital basis (written in second quantization):  
\be\label{eq:unit_transf_2nde_quantization}
\hat{c}_i\rightarrow \sum^{{L}_{\rm tot}}_{j=1} U_{ji}\hat{c}_j.
\ee
Using the block structures of $\bgam$ and ${\bm{U}}_{ee}$ [see Eqs.~(\ref{eq:1RDM_local_basis}) and (\ref{eq:decomp_Uee_unitary_transf})] gives the more explicit expression
\be\label{eq:unit_trans_1RDM_more_explicit}
{\bm U}^\dagger\bgam{\bm U}=
\left[
\begin{matrix}
 \bgam_{ff} & \bgam_{ef}^\dagger{\bm{U}}_{ee} \\
 {\bm{U}}_{ee}^\dagger\bgam_{ef} & {\bm{U}}_{ee}^\dagger\bgam_{ee}{\bm{U}}_{ee}
 \end{matrix}
\right],
\ee
where
\be\label{eq:cenv_f_block_zero}
{\bm{U}}_{ee}^\dagger\bgam_{ef}=
\left[
\begin{matrix}
 {\bm{U}}_{eb}^\dagger \bgam_{ef}  
 \\
 {\bm{U}}_{e\mathcal{E}}^\dagger \bgam_{ef}  
 \end{matrix}
\right]
=
\left[
\begin{matrix}
 {\bm{U}}_{eb}^\dagger \bgam_{ef}  
 \\
{\bm 0}_{\mathcal{E}f}  
 \end{matrix}
\right]
,
\ee
according to Eq.~(\ref{eq:orthogonality_constraint_mathcalE}), thus leading to the following block structure of the density matrix in the embedding representation [see Eq.~(\ref{eq:decomp_fragment_bath_cenv})]: 
\be\label{eq:SVD_transformed_1RDM_block_structure}
{\bm U}^\dagger\bgam{\bm U}=\left[
\begin{matrix}
\bgam_{ff} &  \bgam_{{b}f}^\dagger& 
{\bm 0}_{f\mathcal{E}} 
\\
\bgam_{{b}f}&
\bgam_{{b}{b}}

& 
\bgam_{{\mathcal{E}}{b}}^\dagger
\\
{\bm 0}_{\mathcal{E}f} 
& 
\bgam_{{\mathcal{E}}{b}}
& \bgam_{\mathcal{E}\mathcal{E}}
\end{matrix}
\right],
\ee
where 
\be\label{eq:gamma_bf_expression}
\bgam_{{b}f}={\bm{U}}_{eb}^\dagger \bgam_{ef}
\ee
and
\be
\left[
\begin{matrix}
\bgam_{{b}{b}}
& 
\bgam_{{\mathcal{E}}{b}}^\dagger
\\
\bgam_{{\mathcal{E}}{b}}
& \bgam_{\mathcal{E}\mathcal{E}}
\end{matrix}
\right]={\bm{U}}_{ee}^\dagger\bgam_{ee}{\bm{U}}_{ee}.
\ee
As readily seen from Eqs.~(\ref{eq:unit_trans_1RDM_more_explicit}), (\ref{eq:cenv_f_block_zero}), and (\ref{eq:SVD_transformed_1RDM_block_structure}), the orthogonality constraint of Eq.~(\ref{eq:orthogonality_constraint_mathcalE}) ensures that the fragment is entangled (within the density matrix) only with the bath subspace. The latter is in fact implicitly defined as the orthogonal space to the cluster's environment, which is fully determined from the orthogonality constraint of Eq.~(\ref{eq:orthogonality_constraint_mathcalE}). This more general definition of the bath, which will enable us to connect later on the SVD to the Householder transformation (see Sec.~\ref{sec:Householder}), leaves the possibility to arbitrarily choose the orthonormal bath spin-orbital basis. The SVD-based $\left\{\ket{\phi_b}=\hat{\phi}_b^\dagger\ket{\rm vac}\right\}$ bath basis constructed from Eq.~(\ref{eq:SVD_U1matrix}) [see also Eqs.~(\ref{eq:decomp_Uee_unitary_transf}), (\ref{eq:full_unitary_transf_matrix}), and (\ref{eq:unit_transf_2nde_quantization})], where  
\be\label{eq:bath_spin_orbitals_expansion}
\hat{\phi}_b\underset{{L}_{\rm frag}<b\leq 2{L}_{\rm frag}}{=}\sum_e U_{eb}\hat{c}_e,
\ee
is one choice among others. Interestingly, in the particular single-impurity case (${L}_{\rm frag}=1$), $\mathbfcal{U}$ and $\mathbfcal{D}$ reduce to numbers and the single bath spin-orbital can be written explicitly as a simple functional of the density matrix  [see Eqs.~(\ref{eq:diag_pb_SVD}) and (\ref{eq:SVD_U1matrix})]:
\be
\ket{\phi_b}\overset{{L}_{\rm frag}=1}{=}\frac{1}{\sqrt{\sum_e \abs{\gamma_{ef}}^2}}\sum_e \gamma^*_{ef}\ket{\chi_e}. 
\ee

Let us finally note that, according to Eqs.~(\ref{eq:diag_pb_SVD}), (\ref{eq:SVD_U1matrix}), (\ref{eq:gamma_bf_expression}),  
\be\label{eq:gamma_bf_can_be_inverted}
\bgam_{{b}f}\bgam_{{b}f}^\dagger={\bm{U}}_{eb}^\dagger \bgam_{ef}\bgam_{ef}^\dagger{\bm{U}}_{eb}={\mathbfcal{D}}^{-\frac{1}{2}}{\mathbfcal{D}}^2{\mathbfcal{D}}^{-\frac{1}{2}}={\mathbfcal{D}},
\ee
thus leading to [see Eq.~(\ref{eq:non-zero_square_singular_values_matrix})]
\be\label{eq:abs_det_gamma_bf_strictly_positive}
\abs{\det(\bgam_{{b}f})}=\prod^{L_{\rm frag}}_{b=1}\abs{\sigma_b}>0,
\ee
from which we deduce that $\bgam_{{b}f}$ is {\it invertible}. As further discussed in Sec.~\ref{sec:exact_embedding_idempotRDM}, this property leads to a drastic simplification of the embedding when the reference full-size density matrix $\bgam$ is idempotent.


\section{Quantum bath from the block Householder transformation}\label{sec:Householder}
 
 We start from the exact same full-size density matrix $\bgam$ of Eq.~(\ref{eq:1RDM_local_basis}). A substantial difference between the SVD reviewed in the previous section and the Householder transformation that is used in Ht-DMFET~\cite{sekaran2021,Yalouz2022quantum_embedding} lies in the fact that the latter transformation exploits
the following subblock structure of the environment-fragment density matrix block~\cite{AML99_Rotella_Block_Householder_transf}:
\be\label{eq:gamma_ef_decomp}
\bgam_{ef}=
\left[\begin{matrix}
\bgam_{e_1f} 
\\
\bgam_{e_2f}
\end{matrix}
\right],
\ee  
where the {\it square} matrix 
\be
\bgam_{e_1f}\equiv
\left\{\gamma_{ij}\right\}_{L_{\rm frag}
< i\leq 2L_{\rm frag},1\leq j\leq L_{\rm frag}}
\ee
is assumed to be {\it invertible} and $e_1$ denotes an orthonormal spin-orbital
space of dimension $L_{e_1}=L_{\rm frag}$ with which the fragment is entangled.
The rest of the fragment's environment, which is of dimension $L_{e_2}=L_{\rm tot}-2L_{\rm frag}$ and is
orthogonal to both $f$ and $e_1$ spin-orbital subspaces, is
denoted ${e}_2$. There is obviously some
arbitrariness in the choice of $e_1$ or, equivalently, in the
numbering of the spin-orbitals in the environment. We will see that the choice
of $e_1$ has no impact on the construction of the bath
spin-orbital space, as long as $\bgam^{-1}_{e_1f}$ exists.\\

The general (so-called ``block'', when multiple impurities are embedded~\cite{Yalouz2022quantum_embedding}) Householder transformation that
applies to the full one-electron Hilbert space is represented as follows in the
lattice basis~\cite{AML99_Rotella_Block_Householder_transf},
\be\label{eq:HHtrans_matrix_def}
{\bm R}
={\bm
1}-2\bV\left(\bV^\dagger\bV\right)^{-1}\bV^\dagger,
\ee
where ${\bm
1}$ is the $L_{\rm tot}\times L_{\rm tot}$ identity matrix and
\be\label{eq:construction_V_matrix}
\bV=
\left[
\begin{matrix}
{\bm 0}_{ff}
\\
\left({\bm
1}_{ff}+\tilde{\mathbfcal{U}}\tilde{\mathbfcal{D}}^{\frac{1}{2}}\tilde{\mathbfcal{U}}^\dagger\right)
\bgam_{e_1f}
\\
\bgam_{{e}_2f}
\end{matrix}
\right].
\ee 
The $L_{\rm frag}\times L_{\rm frag}$ unitary $\tilde{\mathbfcal{U}}$ and
diagonal 
\be\label{eq:diag_matrix_tildeD_HH_transf}
\tilde{\mathbfcal{D}}\equiv{\rm
diag}\left\{{d_i}\right\}_{1\leq
i\leq L_{\rm frag}}
\ee
matrices are determined from the following diagonalization problem,
\be\label{diag_prob_HH}
{\bm
1}_{ff}+\left[\bgam_{{e}_2f}\bgam^{-1}_{e_1f}\right]^\dagger\bgam_{{e}_2f}\bgam^{-1}_{e_1f}=\tilde{\mathbfcal{U}}\tilde{\mathbfcal{D}}\tilde{\mathbfcal{U}}^\dagger,          
\ee
where, as readily seen, 
\be\label{eq:eigenvalues_tilde_di_larger_than_one}
{d}_i{\geq} 1,\hspace{0.2cm} \forall\, 1\leq i\leq L_{\rm frag}.
\ee
As a result, the
determinant of $\tilde{\mathbfcal{D}}$ is strictly positive and both
\be
\tilde{\mathbfcal{D}}^{\frac{1}{2}}\equiv{\rm
diag}\left\{\sqrt{{d}_i}\right\}_{1\leq
i\leq L_{\rm frag}}
\ee
and 
$\tilde{\mathbfcal{D}}^{-1}\equiv{\rm diag}\left\{1/{d}_i\right\}_{1\leq
i\leq L_{\rm frag}}$ are well defined. It is interesting to note at this point the similarities and differences between the diagonalization problem of Eq.~(\ref{diag_prob_HH}), which can be rewritten as follows, according to Eq.~(\ref{eq:gamma_ef_decomp}), 
\be\label{eq:diag_prob_HH_comp_with_SVD}
\tilde{\mathbfcal{U}}^\dagger\left(\bgam_{{e}f}\bgam_{e_1f}^{-1}\right)^\dagger\bgam_{{e}f}\bgam_{e_1f}^{-1}\tilde{\mathbfcal{U}}=\tilde{\mathbfcal{D}},
\ee
and the one used in the SVD approach [see Eq.~(\ref{eq:diag_pb_SVD})], where $\bgam_{{e}f}^\dagger\bgam_{{e}f}$ is diagonalized instead.
Note also that the to-be-inverted $\bV^\dagger\bV$ matrix in Eq.~(\ref{eq:HHtrans_matrix_def}), which reads [see Eq.~(\ref{eq:construction_V_matrix})] 
\be
\bV^\dagger\bV&=&
\bgam_{e_1f}^\dagger\left({\bm
1}_{ff}+\tilde{\mathbfcal{U}}\tilde{\mathbfcal{D}}^{\frac{1}{2}}\tilde{\mathbfcal{U}}^\dagger\right)
\left({\bm
1}_{ff}+\tilde{\mathbfcal{U}}\tilde{\mathbfcal{D}}^{\frac{1}{2}}\tilde{\mathbfcal{U}}^\dagger\right)
\bgam_{e_1f}
\nonumber
\\
&&
+\bgam_{{e}_2f}^\dagger\bgam_{{e}_2f},
\ee
can be simplified as follows, according to Eq.~(\ref{diag_prob_HH}),
\be\label{eq:VdaggerV_exp1}
\begin{split}
\bV^\dagger\bV
=2&\Big(
\bgam_{{e}_1f}^\dagger\bgam_{{e}_1f}
+
\bgam_{{e}_2f}^\dagger\bgam_{{e}_2f}
\\
&\quad+\bgam_{e_1f}^\dagger\tilde{\mathbfcal{U}}\tilde{\mathbfcal{D}}^{\frac{1}{2}}\tilde{\mathbfcal{U}}^\dagger\bgam_{e_1f}
\Big),
\end{split}
\ee
or, equivalently,
\be
\bV^\dagger\bV=2\bgam_{e_1f}^\dagger\tilde{\mathbfcal{U}}\tilde{\mathbfcal{D}}^{\frac{1}{2}}\left(\tilde{\mathbfcal{D}}^{\frac{1}{2}}+{\bm
1}_{ff}\right)\tilde{\mathbfcal{U}}^\dagger\bgam_{e_1f},
\ee
from which we immediately deduce [see Eqs.~(\ref{eq:diag_matrix_tildeD_HH_transf}) and (\ref{eq:eigenvalues_tilde_di_larger_than_one})] that the assumed invertibility of $\bgam_{e_1f}$ implies
that of  $\bV^\dagger\bV$. Let us finally emphasize that the Householder transformation
is unitary and Hermitian, by construction [see Eq.~(\ref{eq:HHtrans_matrix_def})]. From a geometrical point of view, it should be
seen as a reflection~\cite{sekaran2021} (not a rotation) that transforms the $L_{\rm frag}$ column vectors collected in $\bV$ into their opposite vectors:
\be
{\bm R}\bV=-\bV.
\ee

In order to compare the Householder transformation to the SVD 
of Sec.~\ref{sec:SVD_EF_block} [see Eqs.~(\ref{eq:decomp_Uee_unitary_transf}) and (\ref{eq:full_unitary_transf_matrix})], we propose to rewrite its matrix representation as follows,
\be\label{eq:block_structure_HH_transf_matrix}
{\bm R}
=
\left[
\begin{matrix}
{\bm 1}_{ff} & {\bm 0}_{f\tilde{b}} & {\bm 0}_{f\tilde{\mathcal{E}}}
\\
{\bm 0}_{ef} & {\bm U}_{e\tilde{b}} & {\bm U}_{e\tilde{\mathcal{E}}}
\end{matrix}
\right],
\ee 
where, according to Eqs.~(\ref{eq:HHtrans_matrix_def}) and
(\ref{eq:construction_V_matrix}), the $(L_{\rm tot}-L_{\rm frag})\times L_{\rm frag}$ block ${\bm U}_{e\tilde{b}}$
and the $(L_{\rm tot}-L_{\rm frag})\times(L_{\rm tot}-2L_{\rm frag})\times$ block ${\bm U}_{e\tilde{\mathcal{E}}}$ read 
\begin{widetext}
\be
{\bm U}_{e\tilde{b}}=
\left[
\begin{matrix}
{\bm 1}_{ff}-2\left({\bm
1}_{ff}+\tilde{\mathbfcal{U}}\tilde{\mathbfcal{D}}^{\frac{1}{2}}\tilde{\mathbfcal{U}}^\dagger\right)
\bgam_{e_1f} \left(\bV^\dagger\bV\right)^{-1}\bgam_{e_1f}^\dagger\left({\bm
1}_{ff}+\tilde{\mathbfcal{U}}\tilde{\mathbfcal{D}}^{\frac{1}{2}}\tilde{\mathbfcal{U}}^\dagger\right)\\
-2\bgam_{{e}_2f}\left(\bV^\dagger\bV\right)^{-1}\bgam_{e_1f}^\dagger\left({\bm
1}_{ff}+\tilde{\mathbfcal{U}}\tilde{\mathbfcal{D}}^{\frac{1}{2}}\tilde{\mathbfcal{U}}^\dagger\right)
\end{matrix}
\right]
\ee
and
\be\label{eq:U2_tilde_matrix}
{\bm U}_{e\tilde{\mathcal{E}}}=
\left[
\begin{matrix}
-2\left({\bm
1}_{ff}+\tilde{\mathbfcal{U}}\tilde{\mathbfcal{D}}^{\frac{1}{2}}\tilde{\mathbfcal{U}}^\dagger\right)
\bgam_{e_1f}\left(\bV^\dagger\bV\right)^{-1}\bgam_{{e}_2f}^\dagger
\\
{\bm
1}_{\tilde{\mathcal{E}}\tilde{\mathcal{E}}}-2\bgam_{{e}_2f}\left(\bV^\dagger\bV\right)^{-1}\bgam_{{e}_2f}^\dagger
\end{matrix}
\right],
\ee
\end{widetext}
respectively. By analogy with Eqs.~(\ref{eq:decomp_Uee_unitary_transf}), (\ref{eq:full_unitary_transf_matrix}), and (\ref{eq:unit_trans_1RDM_more_explicit}), we conclude that the Householder-transformed density matrix has the following block structure, 
\be
{\bm R}^\dagger\bgam {\bm R}
=
\left[
\begin{matrix}
\bgam_{ff} &  \bgam_{\tilde{b}f}^\dagger& 
\bgam^\dagger_{ef}{\bm U}_{e\tilde{\mathcal{E}}}
\\
\bgam_{\tilde{b}f}&
\bgam_{\tilde{b}\tilde{b}}
& 
\bgam_{\tilde{\mathcal{E}}\tilde{b}}^\dagger
\\
{\bm U}_{e\tilde{\mathcal{E}}}^\dagger\bgam_{ef}  
& 
\bgam_{\tilde{\mathcal{E}}\tilde{b}}
& \bgam_{\tilde{\mathcal{E}}\tilde{\mathcal{E}}}
\end{matrix}
\right],
\ee
where
\be\label{eq:bf_dm_block_HH}
\bgam_{\tilde{b}f}&=&{\bm U}_{e\tilde{b}}^\dagger\bgam_{ef}
\ee
and
\be
\bgam_{\tilde{b}\tilde{b}}&=&{\bm U}_{e\tilde{b}}^\dagger\bgam_{ee}{\bm U}_{e\tilde{b}},
\\
\bgam_{\tilde{\mathcal{E}}\tilde{b}}&=&{\bm U}_{e\tilde{\mathcal{E}}}^\dagger\bgam_{ee}{\bm U}_{e\tilde{b}},
\\
\bgam_{\tilde{\mathcal{E}}\tilde{\mathcal{E}}}&=&{\bm U}_{e\tilde{\mathcal{E}}}^\dagger\bgam_{ee}{\bm U}_{e\tilde{\mathcal{E}}}.
\ee
Since, according to Eqs.~(\ref{eq:gamma_ef_decomp}) and (\ref{eq:U2_tilde_matrix}),
\be
\begin{split}
\bgam^\dagger_{ef}{\bm U}_{e\tilde{\mathcal{E}}}&=-2\Big[\bgam_{e_1f}^\dagger
\left({\bm
1}_{ff}+\tilde{\mathbfcal{U}}\tilde{\mathbfcal{D}}^{\frac{1}{2}}\tilde{\mathbfcal{U}}^\dagger\right)\bgam_{e_1f}
\\
&\quad \times \left(\bV^\dagger\bV\right)^{-1}\bgam_{{e}_2f}^\dagger\Big]
\\
&\quad+\bgam_{e_2f}^\dagger-2\bgam_{e_2f}^\dagger\bgam_{{e}_2f}\left(\bV^\dagger\bV\right)^{-1}\bgam_{{e}_2f}^\dagger
\end{split}
\ee
or, equivalently, 
\be
\begin{split}
\bgam^\dagger_{ef}{\bm U}_{e\tilde{\mathcal{E}}}=&\Bigg(
\bV^\dagger\bV-2\bgam_{e_1f}^\dagger
\left({\bm
1}_{ff}+\tilde{\mathbfcal{U}}\tilde{\mathbfcal{D}}^{\frac{1}{2}}\tilde{\mathbfcal{U}}^\dagger\right)\bgam_{e_1f}
\\
&\quad-2\bgam_{e_2f}^\dagger\bgam_{{e}_2f}\Bigg)\left(\bV^\dagger\bV\right)^{-1}\bgam_{{e}_2f}^\dagger,
\end{split}
\ee
we finally conclude from Eq.~(\ref{eq:VdaggerV_exp1}) that
\be\label{eq:no_imp_HHenv_entanglement}
\bgam^\dagger_{ef}{\bm U}_{e\tilde{\mathcal{E}}}={\bm 0}_{f\tilde{\mathcal{E}}}.
\ee
As a result, the Householder-transformed density matrix has a similar block structure to that of the SVD-based unitary transformed density matrix [see Eq.~(\ref{eq:SVD_transformed_1RDM_block_structure})],
\be\label{eq:BHH_transformed_1RDM_block_structure}
{\bm R}^\dagger\bgam {\bm R}
=
\left[
\begin{matrix}
\bgam_{ff} &  \bgam_{\tilde{b}f}^\dagger& 
{\bm 0}_{f\tilde{\mathcal{E}}} 
\\
\bgam_{\tilde{b}f}&
\bgam_{\tilde{b}\tilde{b}}
& 
\bgam_{\tilde{\mathcal{E}}\tilde{b}}^\dagger
\\
{\bm 0}_{\tilde{\mathcal{E}}f} 
& 
\bgam_{\tilde{\mathcal{E}}\tilde{b}}
& \bgam_{\tilde{\mathcal{E}}\tilde{\mathcal{E}}}
\end{matrix}
\right],
\ee
where the embedded fragment is entangled only with the Householder bath, which is defined by ${\bm U}_{e\tilde{b}}$. As further discussed in Sec.~\ref{sec:equivalence_SVD-HH}, the orthogonality relation of Eq.~(\ref{eq:no_imp_HHenv_entanglement}), which echoes the one of Eq.~(\ref{eq:orthogonality_constraint_mathcalE}), is the fundamental reason why SVD and Householder quantum baths are equivalent, even when the reference full-system density matrix $\bgam$ is not idempotent. This generalization of the equivalence in bath between the SVD and the (block) Householder transformation, which has been demonstrated previously in the particular case of idempotent density matrices~\cite{Yalouz2022quantum_embedding}, is the key result of the present work.\\   

Let us finally mention that, like in the SVD-based unitary transformed density matrix [see Eq.~(\ref{eq:abs_det_gamma_bf_strictly_positive})], the bath-fragment block is {\it invertible}, which will be of primary importance when rationalizing the embedding of idempotent density matrices in Sec.~\ref{sec:exact_embedding_idempotRDM}. Indeed, according to Eq.~(\ref{eq:bf_dm_block_HH}) and the orthogonality relation of Eq.~(\ref{eq:no_imp_HHenv_entanglement}), we have    
\be\label{eq:gammabf_dagger_times_gammabf_HH}
\begin{split}
\bgam_{\tilde{b}f}^\dagger\bgam_{\tilde{b}f}&=\bgam_{ef}^\dagger{\bm U}_{e\tilde{b}}{\bm U}_{e\tilde{b}}^\dagger\bgam_{ef}
\\
&=\bgam_{ef}^\dagger\left({\bm U}_{e\tilde{b}}{\bm U}_{e\tilde{b}}^\dagger+{\bm U}_{e\tilde{\mathcal{E}}}{\bm U}_{e\tilde{\mathcal{E}}}^\dagger\right)\bgam_{ef},
\end{split}
\ee
where, as readily seen from Eqs.~(\ref{eq:HHtrans_matrix_def}) and (\ref{eq:block_structure_HH_transf_matrix}), 
\be\label{eq:RI_HH}
{\bm U}_{e\tilde{b}}{\bm U}_{e\tilde{b}}^\dagger+{\bm U}_{e\tilde{\mathcal{E}}}{\bm U}_{e\tilde{\mathcal{E}}}^\dagger=\left[{\bm R}{\bm R}^\dagger\right]_{ee}={\bm 1}_{ee},
\ee
thus leading to [see Eqs.~(\ref{eq:gamma_ef_decomp}) and (\ref{diag_prob_HH})]
\be
\begin{split}
\bgam_{\tilde{b}f}^\dagger\bgam_{\tilde{b}f}
&=\bgam_{ef}^\dagger\bgam_{ef}
=\bgam_{e_1f}^\dagger\bgam_{e_1f}+\bgam_{e_2f}^\dagger\bgam_{e_2f}
\\
&=\bgam_{e_1f}^\dagger\tilde{\mathbfcal{U}}\tilde{\mathbfcal{D}}\tilde{\mathbfcal{U}}^\dagger\bgam_{e_1f},
\end{split}
\ee
and, consequently,
\be\label{eq:abs_det_gammabf_HH_strictly_positive}
\abs{\det(\bgam_{\tilde{b}f})}=\abs{\det(\bgam_{e_1f})}\prod^{L_{\rm frag}}_{i=1}\sqrt{{d}_i}>0.
\ee


\section{Equivalence of SVD and Householder quantum baths}\label{sec:equivalence_SVD-HH}

We prove in this section the equivalence of SVD and Householder quantum baths in the general case where the reference full-system density matrix is not necessarily idempotent. For that purpose, we first need to verify that the Householder bath basis is orthogonal to the SVD-based cluster's environment one. Indeed, according to Eqs.~(\ref{eq:bf_dm_block_HH}), (\ref{eq:no_imp_HHenv_entanglement}), and (\ref{eq:abs_det_gammabf_HH_strictly_positive}),
\be
\begin{split}
{\bm U}_{e\tilde{b}}^\dagger{\bm U}_{e\mathcal{E}}&= \left(\bgam_{\tilde{b}f}^{-1}\right)^\dagger\bgam_{\tilde{b}f}^\dagger {\bm U}_{e\tilde{b}}^\dagger{\bm U}_{e\mathcal{E}}
\\
&=\left(\bgam_{\tilde{b}f}^{-1}\right)^\dagger \bgam_{ef}^\dagger {\bm U}_{e\tilde{b}} {\bm U}_{e\tilde{b}}^\dagger{\bm U}_{e\mathcal{E}}
\\
&= \left(\bgam_{\tilde{b}f}^{-1}\right)^\dagger \bgam_{ef}^\dagger\left({\bm U}_{e\tilde{\mathcal{E}}}{\bm U}_{e\tilde{\mathcal{E}}}^\dagger+{\bm U}_{e\tilde{b}} {\bm U}_{e\tilde{b}}^\dagger\right){\bm U}_{e\mathcal{E}}  
,
\end{split}
\ee
thus leading to [see Eqs.~(\ref{eq:orthogonality_constraint_mathcalE}) and (\ref{eq:RI_HH})]
\be\label{eq:HHbath_ortho_SVD_cluster_env}
{\bm U}_{e\tilde{b}}^\dagger{\bm U}_{e\mathcal{E}}=\left(\bgam_{\tilde{b}f}^{-1}\right)^\dagger \bgam_{ef}^\dagger {\bm U}_{e\mathcal{E}}={\bm 0}_{\tilde{b}\mathcal{E}}.
\ee
If we now expand the Householder bath in the spin-orbital basis generated by the SVD of $\bgam_{ef}$ [see Eq.~(\ref{eq:RI_SVD})],
 \be
{\bm U}_{e\tilde{b}}={\bm 1}_{ee}{\bm U}_{e\tilde{b}}=\left({\bm U}_{eb}{\bm U}_{eb}^\dagger+{\bm U}_{e\mathcal{E}}{\bm U}_{e\mathcal{E}}^\dagger\right){\bm U}_{e\tilde{b}},
\ee
it comes from Eq.~(\ref{eq:HHbath_ortho_SVD_cluster_env}) that
\be\label{eq:unitary_transf_HH_SVD_bath}
{\bm U}_{e\tilde{b}}= {\bm U}_{eb}\mathbfcal{W},
\ee
where the $L_{\rm frag}\times L_{\rm frag}$ overlap matrix between the SVD and Householder bath spin-orbitals 
\be \label{eq:W_matrix}
\mathbfcal{W}={\bm U}_{eb}^\dagger{\bm U}_{e\tilde{b}}
\ee
is, like the Householder transformation [see Eq.~(\ref{eq:block_structure_HH_transf_matrix})], {unitary}:
\be\label{eq:overlap_SVD_HH_unitary}
\begin{split}
\mathbfcal{W}^\dagger\mathbfcal{W}&={\bm U}_{e\tilde{b}}^\dagger{\bm U}_{eb}{\bm U}_{eb}^\dagger{\bm U}_{e\tilde{b}}
\\
&={\bm U}_{e\tilde{b}}^\dagger\left({\bm U}_{e\mathcal{E}}{\bm U}_{e\mathcal{E}}^\dagger+{\bm U}_{eb}{\bm U}_{eb}^\dagger\right){\bm U}_{e\tilde{b}}
\\
&={\bm U}_{e\tilde{b}}^\dagger{\bm U}_{e\tilde{b}}
\\
&=\left[{\bm R}^\dagger{\bm R}\right]_{\tilde{b}\tilde{b}}
\\
&={\bm 1}_{\tilde{b}\tilde{b}}.
\end{split}
\ee
Thus we conclude that the SVD and Householder quantum baths correspond to the {\it same} spin-orbital subspace $\mathcal{B}$ for which, according to Eqs.~(\ref{eq:SVD_U1matrix}) and (\ref{eq:bath_spin_orbitals_expansion}), a trivial (but non-orthonormal) density-matrix functional basis reads
\be\label{eq:simple_expression_1RDM_func_bath}
\mathcal{B}\equiv \mathcal{B}[\bgam]= \left\{\sum_{e}\gamma_{ef}\ket{\chi_e}\right\}_{1\leq f\leq L_{\rm frag}}.
\ee
In other words, as readily seen from Eq.~(\ref{eq:unitary_transf_HH_SVD_bath}), the Householder bath spin-orbitals can be recovered from the SVD ones through a unitary transformation (within the $\mathcal{B}$ subspace).

\section{Density matrix functional exactification of DMET}\label{sec:exactification_dmet}

We briefly revisit in Sec.~\ref{sec:exact_embedding_idempotRDM}, in the light of the previous sections, the well-known embedding of idempotent density matrices and then discuss its extension to correlated density matrices in Sec.~\ref{sec:exact_interacting_embedding}. 

\subsection{The non-interacting or mean-field case}\label{sec:exact_embedding_idempotRDM}

In this section we focus on the most common practical situation where the full-size system is described at the non-interacting or mean-field levels of approximation, so that the density matrix becomes {\it idempotent}. We first consider the SVD-based construction of the bath that is described in Sec.~\ref{sec:SVD_EF_block}. As expected from Sec.~\ref{sec:equivalence_SVD-HH}, the exact same simplifications of the embedding will occur if the Householder transformation is used instead.\\

Starting from the general block structure of the unitary-transformed density matrix in Eq.~(\ref{eq:SVD_transformed_1RDM_block_structure}), we deduce from the (additional) idempotency constraint $\bgam^2=\bgam$ or, equivalently,
\be
\left({\bm U}^\dagger\bgam{\bm U}\right)^2={\bm U}^\dagger\bgam{\bm U},
\ee
that 
\be\label{eq:Ef_block_gamma2}
\bgam_{{\mathcal{E}}{b}}\bgam_{{b}f}={\bm 0}_{\mathcal{E}f}
\ee
and
\be\label{eq:bf_block_gamma2}
\bgam_{{b}f}\bgam_{ff}+\bgam_{{b}{b}}\bgam_{{b}f}=\bgam_{{b}f},
\ee
by considering the cluster's environment-fragment and bath-fragment blocks, respectively. 
Since $\bgam_{{b}f}$ is invertible [see Eq.~(\ref{eq:abs_det_gamma_bf_strictly_positive})], we conclude from Eq.~(\ref{eq:Ef_block_gamma2}) that $\bgam_{{\mathcal{E}}{b}}={\bm 0}_{{\mathcal{E}}{b}}$, which leads to the following block-diagonal structure of the unitary-transformed density matrix [see Eq.~(\ref{eq:SVD_transformed_1RDM_block_structure})]: 
\be\label{eq:SVD_transformed_1RDM_diagonal_block_structure}
{\bm U}^\dagger\bgam{\bm U}=\left[
\begin{matrix}
\bgam_{ff} &  \bgam_{{b}f}^\dagger& 
{\bm 0}_{f\mathcal{E}} 
\\
\bgam_{{b}f}&
\bgam_{{b}{b}}
& 
{\bm 0}_{b{\mathcal{E}}}
\\
{\bm 0}_{\mathcal{E}f} 
& 
{\bm 0}_{{\mathcal{E}}{b}}
& \bgam_{\mathcal{E}\mathcal{E}}
\end{matrix}
\right].
\ee
In addition, it comes from Eq.~(\ref{eq:bf_block_gamma2}) that
\be
\bgam_{ff}+\bgam^{-1}_{{b}f}\bgam_{{b}{b}}\bgam_{{b}f}={\bm 1}_{ff},
\ee
thus leading to the trace equality
\be\label{eq:L_f_electrons_in_the_cluster}
\Tr[\bgam_{ff}]+\Tr[\bgam_{{b}{b}}]={L}_{\rm frag}.
\ee
In summary, as readily seen from Eqs.~(\ref{eq:SVD_transformed_1RDM_diagonal_block_structure}) and (\ref{eq:L_f_electrons_in_the_cluster}), in the particular case where the full system is described with an idempotent density matrix, the embedding cluster is completely disentangled from its environment and it contains exactly ${L}_{\rm frag}$ electrons, which is the number of embedded fragment spin-orbitals.\\

Let us stress that the proof holds if the bath is constructed from the Householder transformation instead, simply because the Householder-transformed density matrix has exactly the same block structure [see Eqs.~(\ref{eq:SVD_transformed_1RDM_block_structure}) and (\ref{eq:BHH_transformed_1RDM_block_structure})] and, in this case, the bath-fragment block is also invertible [see Eq.~(\ref{eq:abs_det_gammabf_HH_strictly_positive})].

\subsection{The interacting case}\label{sec:exact_interacting_embedding}

Unlike the original formulation of DMET~\cite{knizia2012density}, which is based on the Schmidt decomposition of a (possibly correlated) many-electron wave function, the unified density-matrix functional embedding formalism presented in Sec.~\ref{sec:equivalence_SVD-HH} allows for a direct construction of one-electron quantum baths from non-idempotent density matrices [see Eq.~(\ref{eq:simple_expression_1RDM_func_bath})]. As shown in the following, a density matrix functional exactification of DMET can be derived on that basis.\\

Starting from the exact variational expression of the full-size system ground-state energy in {\it one-electron reduced density matrix functional theory} (1RDMFT)~\cite{Gilbert75_Hohenberg-Kohn},
\be\label{eq:VP_1RDMFT}
E=\min_{\bgam}\left\{\sum^{L_{\rm tot}}_{i,j=1}h_{ij}\gamma_{ij}+W[\bgam]\right\},
\ee
where
\be
W[\bgam]=\min_{\Psi\rightarrow \bgam} \langle\Psi\vert\hat{W}\vert\Psi\rangle=\langle\hat{W}\rangle_{\Psi[\bgam]}
\ee
is the universal density matrix functional interaction energy~\cite{levy1979universal}, we consider any fragmentation in the lattice representation of the full second-quantized two-electron repulsion operator:
\be\label{eq:full_frag_decomp_Wee}
\hat{W}=\sum^{{\rm fragments}}_{F}\hat{W}^{F}.
\ee
\rev{Note that, like in practical DMET calculations, the ground states of both the full-size system {\it and} the embedding cluster (see below) are assumed to be {\it pure states}.}
Note also that, for convenience and unlike in conventional DMET calculations~\cite{nusspickel2023effective}, the decomposition in Eq.~(\ref{eq:full_frag_decomp_Wee}) involves fragments that are {\it not} disjoint, simply because we aim at recovering the full two-electron repulsion from the fragmentation. It will become clearer in the following that, if disjoint fragments were considered instead, two-electron interactions between the fragments could then be incorporated into a complementary density matrix functional that would still make the approach formally exact. Returning to Eq.~(\ref{eq:full_frag_decomp_Wee}),       
for each fragment $F$ (of dimension $\dim F=L_F\equiv L_{\rm frag}$), we introduce the following interaction density matrix functional, 
\be
W^{F}[\bgam]=\langle\hat{W}^F\rangle_{\Psi^{L_F}[\bgam]},
\ee
where, in the standard non-interacting bath (NIB) formulation of the embedding~\cite{wouters2016practical}, the $L_F$-electron cluster wave function $\Psi^{L_F}[\bgam]$ fulfills the following ground-state Schr\"{o}dinger equation,
\be\label{eq:cluster_SE_NIB}
\begin{split}
&\left(\hat{\mathcal{P}}\left(\hat{h}+\hat{W}^{F}\right)\hat{\mathcal{P}}
-\sum_{f\in F}\mu_f\hat{c}_f^\dagger\hat{c}_f\right)\ket{\Psi^{L_F}[\bgam]}
\\
&\overset{\rm NIB}{\equiv} \mathscr{E}\ket{\Psi^{L_F}[\bgam]},
\end{split}
\ee
$\hat{h}$ being the one-electron Hamiltonian of the true full-size system [see Eq.~(\ref{eq:full_system_Hamil})]. The operator $\hat{\mathcal{P}}\equiv \hat{P}_{L_F}^{\mathcal{C}^F[\bgam]}$ is the projector onto the $L_F$-electron Fock subspace that is generated from the density-matrix functional one-electron {\it cluster} subspace 
\be
\mathcal{C}^F[\bgam]=F\oplus\mathcal{B}^F[\bgam],
\ee
where the density-matrix functional bath subspace $\mathcal{B}^F[\bgam]$ of fragment $F$, for which an orthonormal basis can be generated either by a SVD or a Householder transformation, is defined in Eq.~(\ref{eq:simple_expression_1RDM_func_bath}). The potentials $\left\{\mu_f\right\}$ that we added on each embedded impurity [see Eq.~(\ref{eq:cluster_SE_NIB})] are adjusted such that the $L_F$-electron ground state of the embedding cluster reproduces the occupations of the fragment spin-orbitals in the full-size system:
\be\label{eq:constraint_occ_embedded_imps}
\langle\hat{c}_f^\dagger\hat{c}_f\rangle_{\Psi^{L_F}[\bgam]}\underset{f\in F}{\overset{!}{=}}\gamma_{ff}.
\ee
At this point we should stress that, at the mean-field level of calculation, the embedding procedure becomes exact because the density matrix is idempotent (see Sec.~\ref{sec:exact_embedding_idempotRDM}). However, the exact embedding cluster, constructed according to Eq.~(\ref{eq:simple_expression_1RDM_func_bath}) from the exact (non-idempotent) ground-state density matrix of the full system, is in principle an {\it open} quantum system~\cite{sekaran2021} whose description by the pure-state wave function $\Psi^{L_F}[\bgam]$ is therefore approximate. As a result, the sum of density matrix functional fragment interaction energies $W^{F}[\bgam]$ must be complemented by a correlation functional, which is usually ignored in practice,
\be\label{eq:exact_frag_decomp_inter_func}
W[\bgam]=\sum^{{\rm fragments}}_{F}W^{F}[\bgam]+\overline{W}_{\rm c}[\bgam],
\ee
in order to recover, in principle variationally, the exact ground-state energy of the full system [see Eq.~(\ref{eq:VP_1RDMFT})]. Using {\it multi-reference perturbation theory} (MRPT) has been envisioned for developing approximations to $\overline{W}_{\rm c}[\bgam]$~\cite{sekaran2021}. Note also that the {\it local} version of the correlation potential, which is usually introduced in the full-size system for optimizing the bath self-consistently~\cite{knizia2012density,bulik2014density}, can be interpreted as a functional of the density in the context of DFT for lattices~\cite{mordovina2019self,sekaran2022local}. From this perspective, practical DET~\cite{bulik2014density} becomes a density-functional approximation~\cite{sekaran2022local}. Establishing a clearer formal connection between the non-local version of the correlation potential and the functional derivative $\partial W[\bgam]/\partial \gamma_{ij}$, along the lines of Ref.~\cite{sekaran2022local}, would be an important step towards the rationalization of conventional (multiple-impurity) self-consistent DMET calculations. Moreover, in order to turn the theory into a reliable (and ideally variational~\cite{Lin2022Variational}) computational method, one should pay attention to the (global) $N$-representability~\cite{nusspickel2023effective} of the density-matrix functional decomposition in Eq.~(\ref{eq:exact_frag_decomp_inter_func}). Work is currently in progress in these directions.\\

\rev{Note finally that the NIB formalism used in Eq.~(\ref{eq:cluster_SE_NIB}) emerges naturally when connecting DMET to DFT~\cite{sekaran2022local}. Indeed, from a density functional perspective, the role of the (chemical) potential that is traditionally introduced into the embedded fragment is to exactly compensate the effects of the two-electron repulsion in the fragment on the fragment density, the embedding being already exact for the non-interacting KS system. Thus, a clear connection between the Hartree-exchange-correlation potential of DFT (on the fragment) and the fragment chemical potential of DMET can be established~\cite{sekaran2022local}. Nevertheless, numerical calculations have shown that introducing interactions into the bath gives, in some cases, more accurate energies. This is the case, for example, in the half-filled one-dimensional Hubbard model with a single embedded  impurity (see Fig. 7 in Ref.~\onlinecite{sekaran2021}). On the other hand, the NIB flavor of DMET turns out to perform slightly better than the interacting bath (IB) one away from half filling (see Fig. 9 in Ref.~\onlinecite{sekaran2021}).} For completeness, let us point out that the exact decomposition of the interaction functional in Eq.~(\ref{eq:exact_frag_decomp_inter_func}) can alternatively be based on the IB formulation of the embedding, where the embedded electrons interact in both the fragment and the bath, under the constraint of Eq.~(\ref{eq:constraint_occ_embedded_imps}), thus leading to the following IB Schr\"{o}dinger equation,
\be\label{eq:cluster_SE_IB}
\begin{split}
&\left(\hat{\mathcal{P}}\hat{H}\hat{\mathcal{P}}
-\sum_{f\in F}\mu_f\hat{c}_f^\dagger\hat{c}_f\right)\ket{\Psi^{L_F}[\bgam]}\overset{\rm IB}{\equiv} \mathscr{E}\ket{\Psi^{L_F}[\bgam]},
\end{split}
\ee
$\hat{H}$ being the Hamiltonian of the true full-size system [see Eq.~(\ref{eq:full_system_Hamil})].


\section{Illustrative examples}\label{sec:numerical_examples}

In this section, we illustrate numerically the equivalence between the one-electron quantum bath generated by applying the SVD to the environment-fragment density matrix block and the one obtained by applying the block Householder transformation to the density matrix. For the latter, we used the already implemented transformation of the QuantNBody package \cite{yalouz2022quantnbody}. Calculations have been performed on simple but nontrivial $L$-site ($L=10$ or $L=100$) homogeneous one-dimensional Hubbard models. In this case, the full-system Hamiltonian of Eq.~(\ref{eq:full_system_Hamil}) is simplified as follows,
\be
\begin{split}
&\hat{H}\rightarrow -t \sum^L_{<p,q>}\sum_{\sigma=\uparrow,\downarrow}\hat{c}_{p\sigma}^\dagger\hat{c}_{q\sigma}+U\sum^L_{p}\hat{c}_{p\uparrow}^\dagger\hat{c}_{p\uparrow}\hat{c}_{p\downarrow}^\dagger\hat{c}_{p\downarrow}, 
\end{split}
\ee
where $t$ (which is set to $t=1$ in the following) and $U$ are the (nearest-neighbor) hopping and on-site two-electron repulsion parameters, respectively. In the following, we denote $N$ the number of electrons in the full-size system and $n=N/L$ its uniform density (filling). Periodic boundary conditions have been used. Let us start with the 10-site model at half-filling ($n=1$) and the particular case of $N_{\rm imp}=3$ embedded orbital impurities (which corresponds to $L_{\rm frag}=6$ embedded spin-orbitals). The overlap matrix between the SVD and Householder bath spin-orbitals [see Eq.~(\ref{eq:W_matrix})] that we first computed from the non-interacting full-size system density matrix reads  
\be \label{eq:W_Nimp3_half_fil}
\mathbfcal{W} \overset{\bgam^2=\bgam}{\equiv} 
\begin{bmatrix}
-0.6479 & -0.6770 & -0.3495 \\
-0.5958 & 0.7360  & -0.3215 \\
0.4749 & 0 & -0.8801
\end{bmatrix},
\ee 
and, as expected from a previous work [see Appendix D in Ref.~\onlinecite{Yalouz2022quantum_embedding}] and Eq.~(\ref{eq:overlap_SVD_HH_unitary}), it is unitary. We now turn to the even more interesting case where the reference density matrix $\bgam$ is correlated. The latter can be obtained, for example, from a NOFT calculation. We used the  Cs\'anyi--Arias (CA) functional~\cite{Csanyi00_Tensor}, still with  $N_{\rm imp} = 3$ and  $n = 1$, as a proof of concept, and obtained the following unitary overlap matrix for $U/t=4$:
\be \label{eq:W_Nimp3_half_fil_CA}
\mathbfcal{W}\overset{\bgam^2\neq\bgam}{\equiv}  
\begin{bmatrix}
0 &	1 &	0 \\
0.7878 & 0 & 0.6159\\
-0.6159 & 0 &	0.7878
\end{bmatrix}.
\ee 
\rev{Let us stress that, in this section, NOFT is only used  to generate a reference full-size correlated density matrix from which the embedding cluster (where the one-electron bath is correlated through that density matrix) is constructed. Ultimately, the correlation energy is computed from the embedding cluster's wave function, not from the natural orbital functional. In this respect, testing a simple functional such as the CA one for generating (at a relatively low cost) a correlated density matrix, even though it is known to have problems in the reconstruction of the two-electron reduced density matrix~\cite{herbert2002comparison,herbert2003n}, is relevant. Of course, as discussed further in the following, we still need to verify if the resulting (forced-to-be-closed) embedding cluster can provide sensible and useful results.} Turning back to the comparison of quantum baths, as readily seen from Eqs.~(\ref{eq:W_Nimp3_half_fil}) and (\ref{eq:W_Nimp3_half_fil_CA}), the SVD and Householder bath spin-orbitals may not be the same, but the SVD bath is simply obtained by applying a unitary transformation {\it within} the set of Householder bath spin-orbitals, thus confirming the equivalence of the two bath subspaces, whether the reference density matrix of the full-size system is idempotent or not.\\

\begin{center}
\begin{figure}[h]
\includegraphics[scale=0.50]{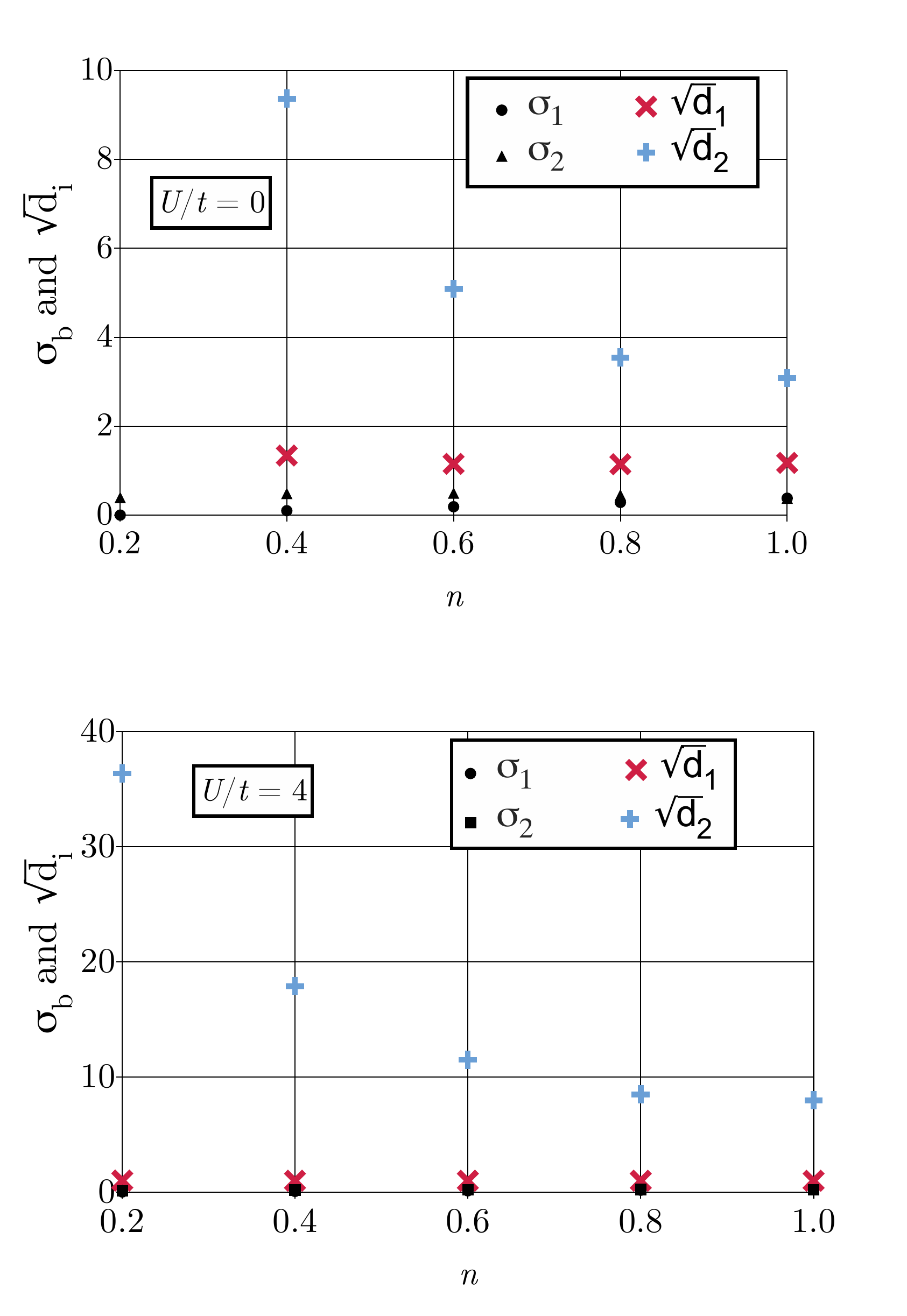}
\caption{Comparison of Householder square root eigenvalues $\left\{\sqrt{d_i}\right\}$ and singular values $\left\{\sigma_b\right\}$ [see Eqs.~(\ref{eq:non-zero_square_singular_values_matrix}), (\ref{eq:diag_pb_SVD}), (\ref{eq:diag_matrix_tildeD_HH_transf}), and (\ref{eq:diag_prob_HH_comp_with_SVD})] computed from non-interacting (top panel) and correlated [using the CA natural orbital functional with $U$/$t$ = 4] (bottom panel) reference density matrices for a $10$-site ring at different fillings $n$ with $N_{\rm imp}=2$ embedded orbital impurities. See text for further details.}
\label{fig1}
\end{figure}
\end{center}

\begin{center}
\begin{figure}[h]
\includegraphics[scale=0.70]{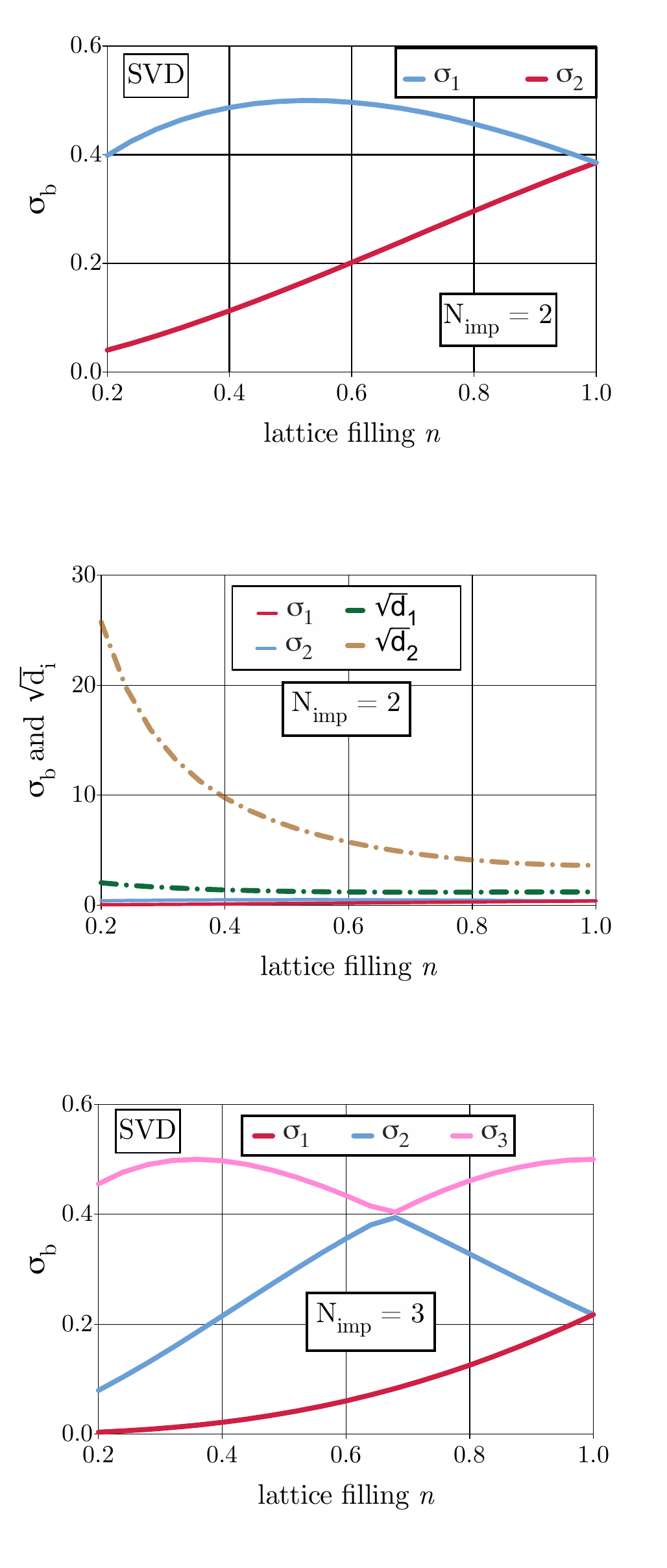}
\caption{Same as Fig.~\ref{fig1} for a non-interacting $100$-site ring. Singular values only are shown in the top panel, for clarity. For comparison, the singular values obtained for $N_{\rm imp}=3$ embedded orbital impurities are shown in the bottom panel.}
\label{fig2}
\end{figure}
\end{center}

For analysis purposes, we show in Fig.~\ref{fig1} the singular values and square root Householder eigenvalues obtained (for $N_{\rm imp}=2$ embedded orbitals and different fillings) from the diagonalization problems that are solved in the SVD and the block Householder transformation, respectively [see Eqs.~(\ref{eq:diag_pb_SVD}) and (\ref{eq:diag_prob_HH_comp_with_SVD})]. Note that the Householder transformation cannot be applied to two non-interacting electrons ($n=0.2$) since the density matrix elements are all identical in this case and, therefore, $\bgam_{e_1f}$ is not invertible. As expected, one of the singular values equals zero in this case (see the top panel of Fig.~\ref{fig1}). 
In connection with these observations, one of the square root Householder eigenvalues becomes significantly larger than both singular values when the filling is low. In order to explore further the low density regime, we applied the same (non-interacting) embedding strategy to a larger $100$-site ring for which the same pattern is observed (see the top and middle panels of Fig.~\ref{fig2}). For completeness, we verified that, even though the embedding of an additional orbital impurity ($N_{\rm imp}=3$) affects both the singular values (see the bottom panel of Fig.~\ref{fig2}) and the square root Householder eigenvalues (not shown), their order of magnitude remains the same. In fact, in all these cases, the reference density matrix $\bgam$ is idempotent, thus leading to $\bgam_{ef}^\dagger\bgam_{ef}=\bgam_{ff}({\bm 1}_{ff}-\bgam_{ff})$, according to Eq.~(\ref{eq:1RDM_local_basis}), where the eigenvalues of $\bgam_{ff}$ are positive and lower than 1. This is the reason why the singular values are in the same range. Note finally that the difference in order of magnitude between the singular values and the square root Householder eigenvalues can be enhanced by electron correlation (see the top and bottom panels of Fig.~\ref{fig1}). In conclusion, if we aim at embedding large fragments and decide, for computational reasons, to reduce the size of the bath by selecting the largest singular values up to a given threshold, one should keep in mind that, when the Householder transformation is employed instead, the order of magnitude of the square root Householder eigenvalues can be substantially different.\\ 

\begin{center}
\begin{figure}[h]
\includegraphics[scale=0.5]{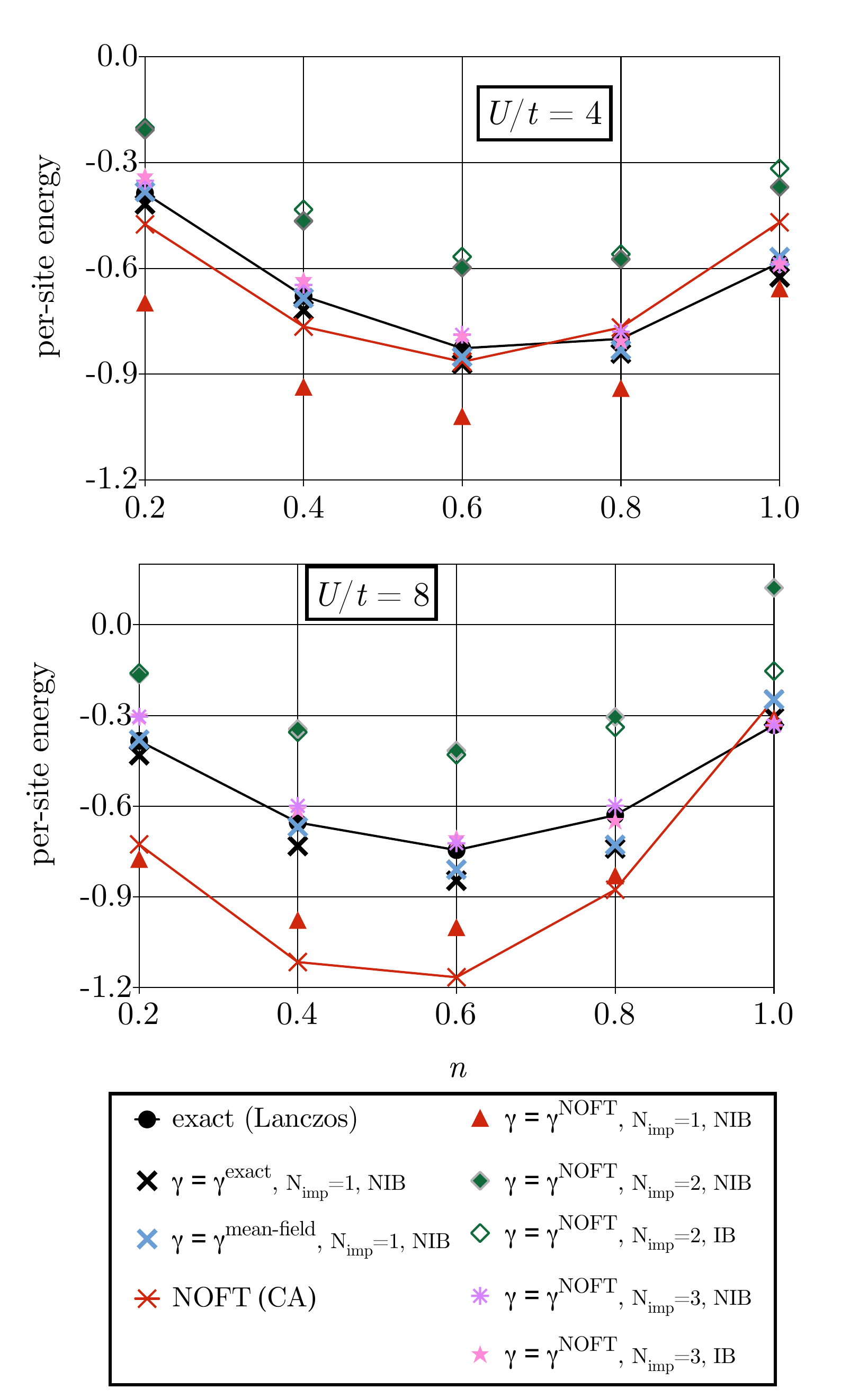}
\caption{Per-site energies obtained for a $10$-site Hubbard ring (as a function of the filling $n$) from the single-shot embedding of $N_{\rm imp}$ orbital impurities ($N_{\rm imp}=1,2,3$) using a correlated density matrix (generated from a NOFT calculation with the CA functional) as reference. Single-impurity per-site energies obtained from the exact and mean-field reference density matrices without interactions in the bath (NIB) are shown for analysis purposes. Comparison is made with exact and NOFT (CA) per-site energies. Results are shown for the challenging strongly correlated $U/t=4$ (top panel) and $U/t=8$ (bottom panel) regimes. See text for further details.}
\label{fig3}
\end{figure}
\end{center}

Let us now discuss the density-matrix functional formulation of DMET proposed in Sec.~\ref{sec:exact_interacting_embedding}. In order to explore the benefit of using a correlated reference density matrix, instead of an idempotent one like in standard DMET implementations, we performed a single-shot embedding of $N_{\rm imp}$ impurities ($N_{\rm imp}=1,2,3$) in a $10$-site Hubbard ring for different fillings. The complementary correlation energy $\overline{W}_{\rm c}[\bgam]$, which has been introduced in Eq.~(\ref{eq:exact_frag_decomp_inter_func}) and for which density matrix functional approximations should be developed, has been neglected. Per-site energies obtained {\it via} the embedding from the exact and approximate reference density matrices are shown in Fig.~\ref{fig3}. Note that both the kinetic (hopping) energy and the on-site repulsion energy were computed from the correlated ground-state wave function of the embedding cluster. Comparison is made with regular NOFT (using the CA functional) for analysis purposes. At the simplest NIB single-impurity level of embedding, using the exact density matrix in place of the mean-field one does not induce a drastic change in the per-site energies, which are relatively close to the exact ones. The fact that the error in energy always increases slightly (except at half-filling when $U/t=8$) when the exact density matrix is employed as reference might be related to the fact that, in this case, the true embedding cluster is not a closed system~\cite{sekaran2021}, unlike in our calculations. In other words, the missing density matrix functional energy contribution $\overline{W}_{\rm c}[\bgam]$ may play an important role in this case. Away from half-filling, the situation is completely different if we use instead the (approximate) correlated CA density matrix as reference. The per-site energies are much too low in this case. We also note that, when $U/t=4$, regular NOFT (CA), which uses the same density matrix, performs much better [see the top panel of Fig.~\ref{fig3}]. This could be a consequence of error cancellations in the computation of the NOFT energies (see Refs.~\onlinecite{Mitxelena2017_on_the_performance,Mitxelena2018_corrigendum} for a more detailed analysis of NOFT calculations in the Hubbard model). On the other hand, in the stronger $U/t=8$ correlation regime, quantum embedding slightly improves on the NOFT (CA) results [see the bottom panel of Fig.~\ref{fig3}]. If we now consider an enlarged embedding cluster with $N_{\rm imp}=2$ orbital impurities (and still a NIB), the per-site energies change drastically and become too high. An unphysical positive energy is even obtained at half-filling when $U/t=8$. Taking into account interactions in the bath (IB scheme) drastically improves on the results when approaching half-filling. This observation raises an important point, namely that there might be some inconsistency in using a NIB within an embedding cluster that has been generated from a full-size density matrix where {\it all} electronic correlations are in principle described. In the present case, the expected overestimation of the impurity site double occupation [see Fig. 8 of Ref.~\onlinecite{sekaran2021}] will be a source of error. Note that accurate results are finally obtained when $N_{\rm imp}=3$ orbitals are embedded, which is a major improvement over the CA functional, especially in the stronger $U/t=8$ correlation regime. In this case, the cluster is large enough such that there is no major difference between NIB and IB embeddings.\\

In conclusion, the embedding cluster can be quite sensitive to the reference full-size density matrix from which it is constructed, especially when a limited number of impurities is embedded. Using better natural orbital functionals~\cite{Mitxelena2017_on_the_performance,Mitxelena2018_corrigendum} in conjunction with an appropriate approximation to $\overline{W}_{\rm c}[\bgam]$ is expected to improve on the convergence of the embedding with respect to the number of impurities. Work is in progress in this direction.  




\section{Conclusions and perspectives}\label{sec:conclusions}

A mathematical proof of the equivalence in one-electron quantum bath between DMET (which uses the SVD) and the more recent Ht-DMFET (which uses the Householder transformation) has been derived in the general case where the reference full-size (one-electron reduced) density matrix is not necessarily idempotent. The two approaches rely on similar but different diagonalization problems whose dimension equals the number of to-be-embedded fragment spin-orbitals (the so-called impurities). A practical advantage of the Householder transformation over the SVD lies in the automatic construction of a complete orthonormal basis for the one-electron Hilbert space, which will become necessary if, ultimately, we aim at describing electron correlation beyond the standard ``closed \rev{embedding} cluster'' approximation. Multi-reference perturbation theory could be used for that purpose (work in progress). One important application of the present unified formulation of quantum embedding theory is the density-matrix functional exactification of DMET. Numerical examples show that the embedding cluster can be quite sensitive to the level of accuracy of the reference full-size density matrix from which it is constructed. The optimal combination of natural orbital functional approximations with quantum embedding, in terms of computational cost and accuracy, should be further explored numerically. This is left for future work.

\section*{Acknowledgements}

The authors would like to thank LabEx CSC (Grant No. ANR-10-LABX-0026-CSC) and the ANR (Grants No. ANR-19-CE07-0024-02 and ANR-19-CE29-0007-01) for funding.
The authors are grateful to Lionel Lacombe, Saad Yalouz, Matthieu Sauban\`{e}re, and Quentin Mar\'{e}cat for fruitful discussions. They also warmly thank Matthieu Sauban\`{e}re for having made his Lanczos code and implementation of the CA natural orbital functional available.



\newcommand{\Aa}[0]{Aa}

\end{document}